\documentclass[a4paper,11pt]{article}
\pdfoutput=1 

\usepackage{jinstpub} 

\usepackage{float}
\usepackage[export]{adjustbox}
\usepackage[utf8]{inputenc}     
\usepackage{textcomp}           
\usepackage{siunitx}            
\usepackage{enumitem}

\title{\boldmath Development of Low-Mass Flex PCB and Nanowire Interconnect Technologies for HEP Module Integration}

\author[a,1]{A. Sharma\note{Corresponding author.},}
\author[b]{A. Gorane,}
\author[a]{J. Weick,}
\author[a]{P. Riedler}

\affiliation[a]{CERN, Switzerland}
\affiliation[b]{University of Freiburg, Germany}

\emailAdd{Abhishek.Sharma@CERN.ch}

\abstract{The development of lightweight flex PCBs and nanowire-based thermal interfaces for low-mass, high-performance detector modules are presented. A novel manufacturing approach targeting flex circuits with double-sided pad access, assembled using ACF and gold studs. Signal integrity was simulated and validation trials conducted on test structures. For thermal management, sintered nanowire interfaces were evaluated. These results contribute quantitative input relevant to minimal-mass, scalable packaging in HEP detectors.}


\proceeding{Topical Workshop on Electronics for Particle Physics\\
  6th - 10th October 2025\\
  Crete, Greece}

\begin{document}
\maketitle
\flushbottom

\section{Introduction}
\label{sec:intro}

Future upgrades of high-energy physics (HEP) detectors demand innovative solutions that reduce the material budget while maintaining reliable electrical performance and efficient thermal management. Lowering the radiation length of detector modules is essential for achieving the precision required at the High-Luminosity LHC~\cite{HL_LHC} and beyond. This must be accomplished without sacrificing robustness or manufacturability.

\begin{figure}[H]
\centering 
\includegraphics[width=15cm]{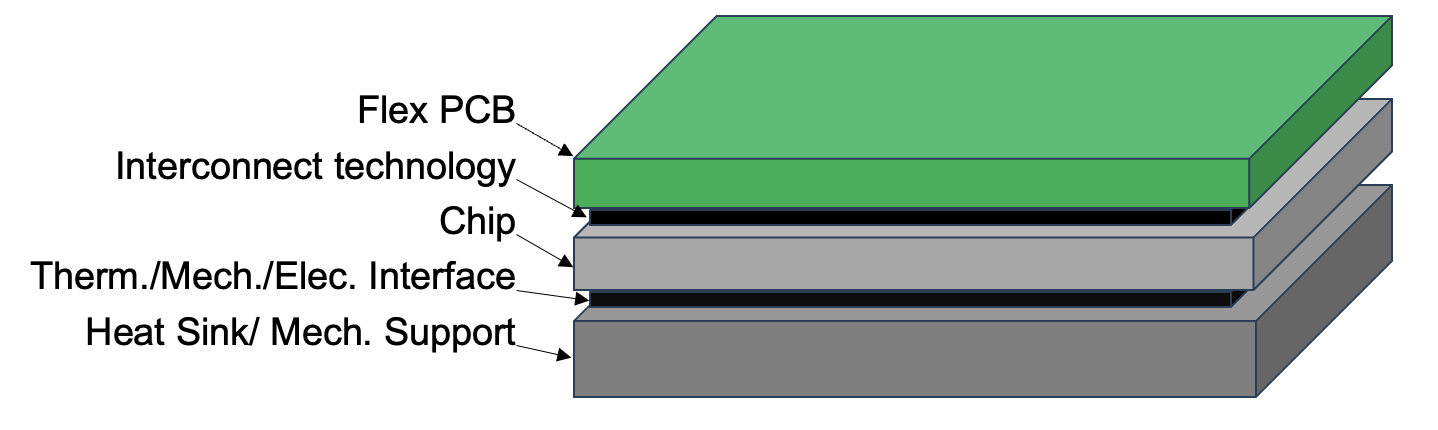}
\caption{\label{fig:LayerStack} Sketch of typical module to mechanical support layer stack.}
\end{figure}

Two complementary developments are being pursued within the scope of this study. The first is the fabrication of ultra-thin flex PCBs that provide double-sided pad access for low-mass interconnections. These structures enable dense routing in compact module geometries while reducing overall weight. The second is the evaluation of nanowire assemblies as novel thermal and electrical interfaces. Their large surface area makes them promising candidates for enhanced heat transfer and electrical conduction, offering possibilities such as back-side biasing of MAPS sensors or partial thermal bonding of ASICs. Figure~\ref{fig:LayerStack} shows the mechanical support layer stack typically required for a module.

Together, these approaches target compact, lightweight, and scalable module assemblies that combine minimised material, improved heat extraction, and robust electrical links for operation in high-rate, high-radiation environments.

\section{Ultra-Thin Flex PCB}
\label{sec:Ultra-Thin Flex PCB}

\begin{figure}[H]
\centering 
\includegraphics[width=15cm]{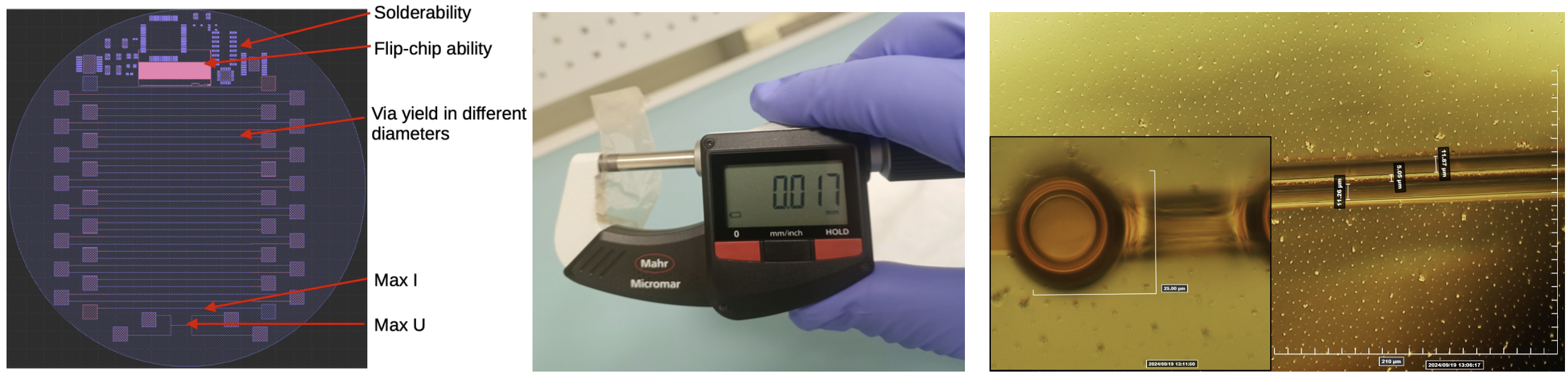}
\caption{\label{fig:Wafer-micrometer-vias-traces} Flex PCB with dedicated test features produced on 4" silicon wafer substrate (left). 17~\textmu m flex thickness measured with micrometer (centre). 25~\textmu m diameter vias and 10~\textmu m traces with 5~\textmu m gaps (right).}
\end{figure}

\begin{figure}[H]
\centering 
\includegraphics[width=15cm]{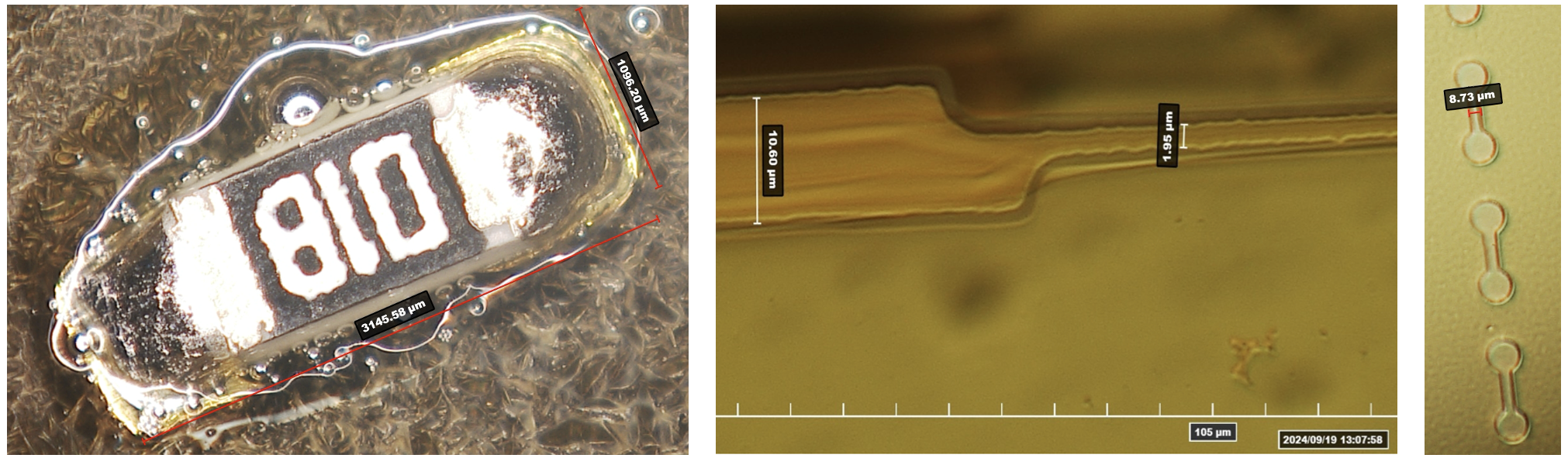}
\caption{\label{fig:Solderability-Fuse-SmallFeatures} Successful solderability of 0603 SMD component on bottom pads (left). Minimum trace width tests including current evaluations on titanium (centre). 8~\textmu m small feature production tests (right).}
\end{figure}

One of the goals of this work is to achieve a flexible, low-material and reliable module packaging approach. The process flow was executed at Campus Biotech~\cite{campusbiotech} and EPFL-CMi~\cite{EPFL_CMi}. A circular flexible circuit was developed on a 4-inch silicon wafer to prototype and validate the process flows. The design incorporates via yield test arrays, regions for assessing the solderability of both top and bottom layers, dedicated flip-chip testing zones, and structures to evaluate the maximum current and voltage handling capabilities, as shown in Fig.~\ref{fig:Wafer-micrometer-vias-traces} and Fig.~\ref{fig:Solderability-Fuse-SmallFeatures}.

The detachment layer of a printed circuit board (PCB) was produced on a silicon wafer substrate. The key objective was to enable forceless separation of the flexible circuit and to explore the potential fabrication of a bottom-pad layer. Iterative optimizations of the base layer stack were performed to achieve a clean and intact separation of the flex from the substrate.

\begin{figure}[H]
\centering 
\includegraphics[width=15cm]{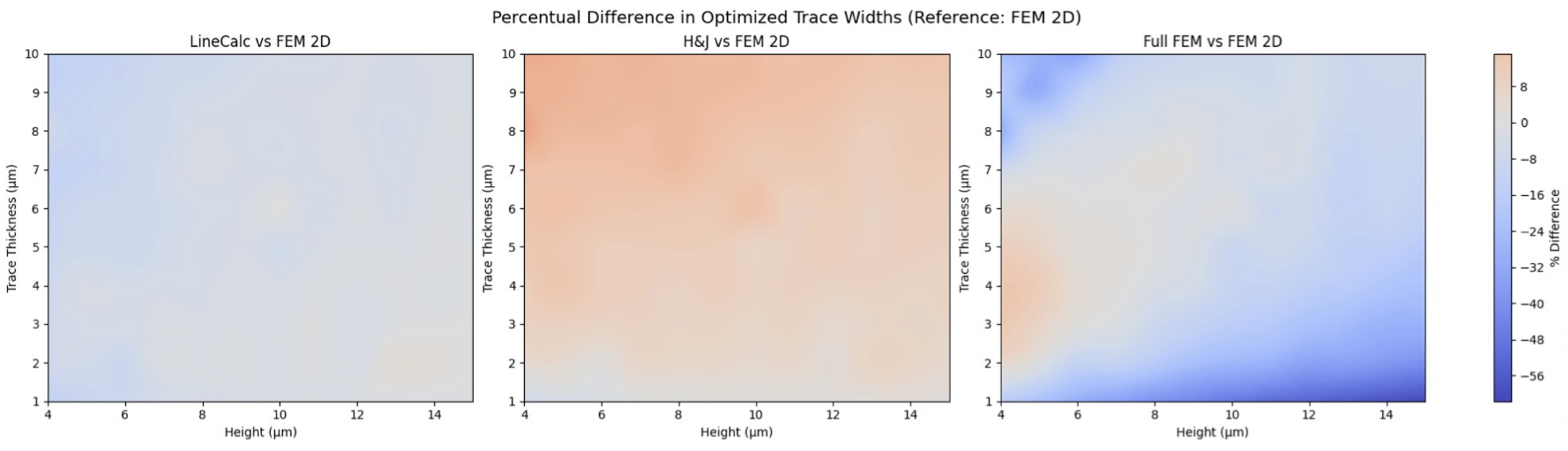}
\caption{\label{fig:Simulations-Differences} Relative difference between obtained widths for Z0 = 50$\Omega$ compared to the 2D FEM method for LineCalc, Hammerstad and Jensen, and Full FEM models.}
\end{figure}

Trace impedance simulation studies were conducted with the aim of simulating small trace geometries (approximately 5-40~\textmu m wide and 2–10~\textmu m thick) and the corresponding layer stack to achieve defined impedance values, shown in Fig.~\ref{fig:Simulations-Differences}.

\begin{figure}[H]
\centering 
\includegraphics[width=7.5cm]{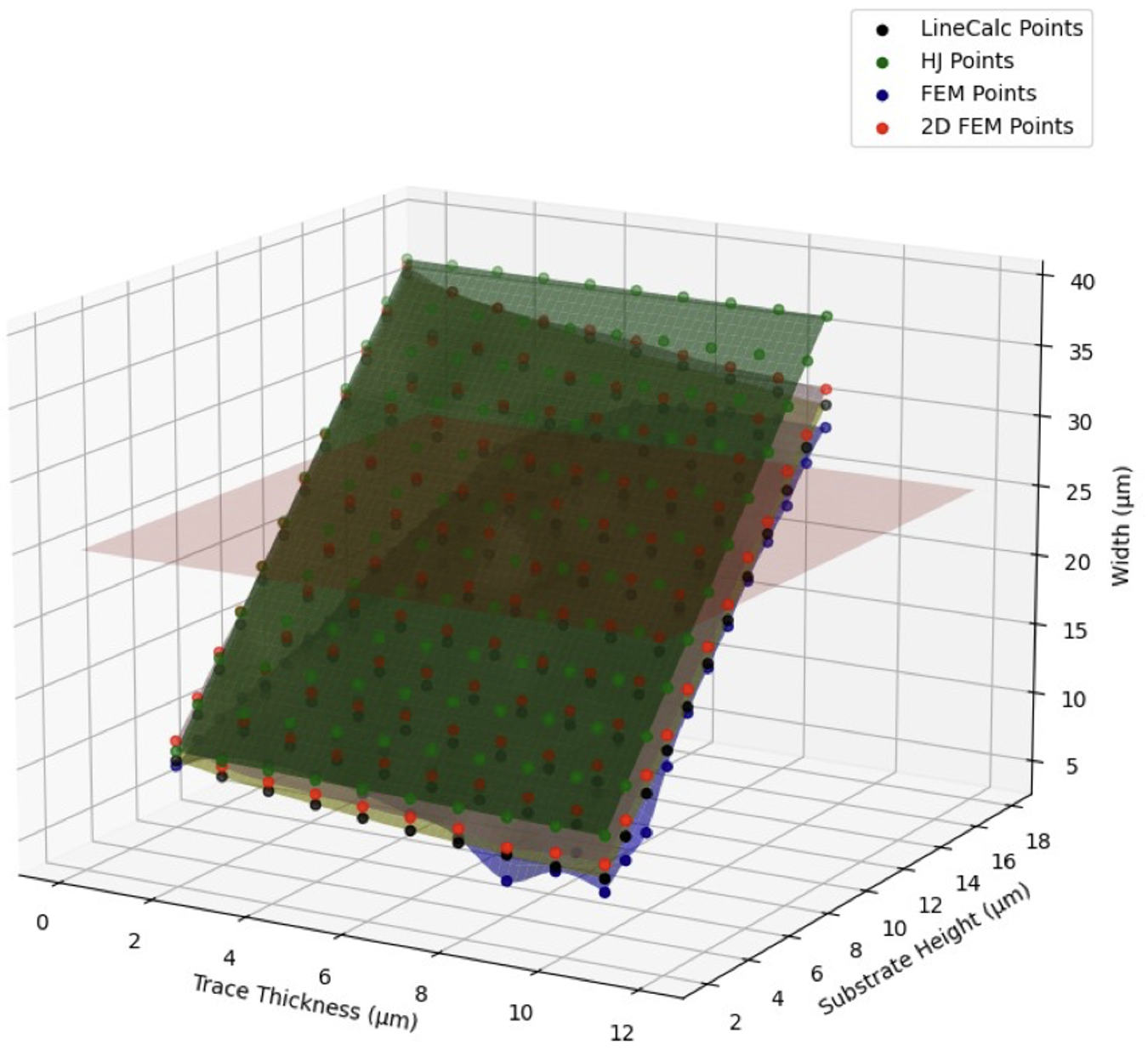}
\caption{\label{fig:Simulations} Trace dimensions obtained for Z0 = 50$\Omega$, for 3D analysis methods.}
\end{figure}

Four different models were compared: Hammerstad and Jensen~\cite{hammerstad_jensen}, ADS Keysight LineCalc~\cite{ads_linecalc}, and Ansys HFSS~\cite{ansys_hfss} in both 2D and 3D configurations. The divergences observed in the 3D simulations were further analyzed and quantified in Fig.~\ref{fig:Simulations} for high trace geometry aspect ratios.

\section{Nano Wires for Pad Interconnection}
\label{sec:Nano Wires for Pad Interconnection}

Nanowire applications were explored in two distinct configurations. In the first, 6 µm nanowires were grown directly on individual pads to enable electrical interconnections. In the second, nanowires were applied on a tape across large areas to facilitate thermal interconnections.

The ongoing investigations focus on evaluating whether nanowires can effectively serve as dual-purpose interconnects, providing both efficient heat transfer and mechanical bonding to a support structure. In particular, the study explores the feasibility of partial backside thermal bonding on ASICs and examines whether this approach could be extended to enable backside voltage interconnections, potentially combining thermal and electrical functionality within the same interface.

Two nanowire bonding methods were examined:
\begin{itemize}[topsep=0pt, noitemsep]
    \item \textbf{Glue-supported bonding} – Nanowires of approximately 1~\textmu m diameter and 10~\textmu m length were used to establish thermal connections, while the adhesive layer provided mechanical stability, achieving bonding process pressures of around $5~\text{MPa}$ at room temperature.
    \item \textbf{Sintering-based bonding} – In this method, nanowires with 100~\textmu m diameter and 1~\textmu m length were sintered with the opposing pad, resulting in significantly higher bonding process pressures, reaching about $20~\text{MPa}$ at $230~^{\circ}\text{C}$.
\end{itemize}

\begin{figure}[H]
\centering 
\includegraphics[width=15cm]{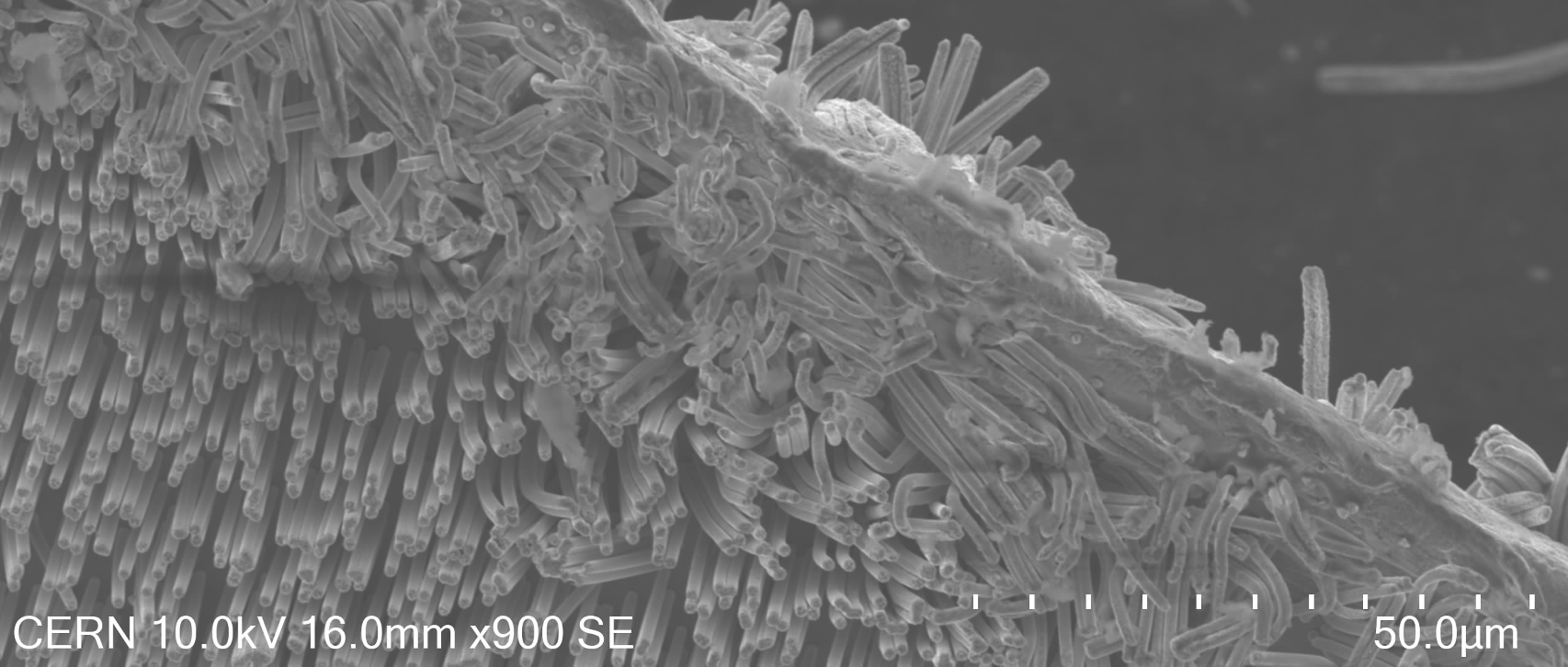}
\caption{\label{fig:Nanowires-tape} Nano-wires on tape.}
\end{figure}

Seven different thermal interconnection technologies are currently being assessed, including nanowire-based approaches (both glue-supported and sintered), silver paste, Araldite (tested under 5 MPa pressure and under forceless conditions), Stycast, and thermal epoxy.

\begin{figure}[H]
\centering 
\includegraphics[width=15cm]{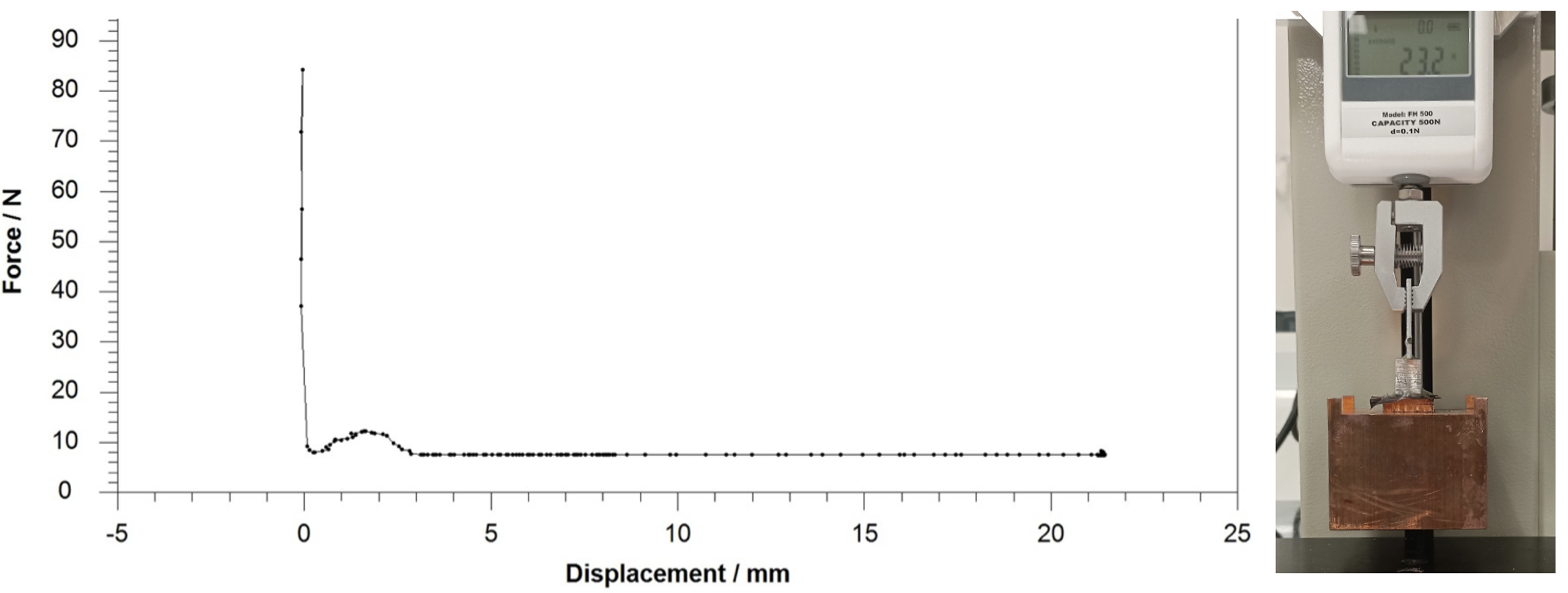}
\caption{\label{fig:PullTestClean} Pull-test measurement of a sintered nano-wire assembly.}
\end{figure}

The initial sintering tests demonstrated successful mechanical properties, indicating the feasibility of this bonding approach. Fig.~\ref{fig:PullTestClean} shows a sintered nano-wire assembly during pull-test able to sustain over 80N of vertical pull force before failure.

\pagebreak
\acknowledgments

The developments presented in this contribution are performed within the DRD3 collaboration on Solid-State Detectors and the CERN EP R\&D programme on technologies for future experiments.

\end{document}